\documentclass[conference]{IEEEtran}

\usepackage{color}
\usepackage{url}
\usepackage{multirow}
\usepackage{rotating}
\usepackage{array}
\usepackage{color}
\usepackage{verbatim}
\usepackage{bigstrut}
\usepackage{multirow}
\usepackage{amsmath}
\usepackage{amssymb}
\usepackage{setspace}
\usepackage{diagbox}
\usepackage{url}
\usepackage{pifont}
\usepackage{subfigure}
\usepackage[ruled,vlined, linesnumbered]{algorithm2e}
\usepackage{graphicx}
\usepackage{caption}
\usepackage{enumitem}
\usepackage{xcolor}
\usepackage{balance}
\newcolumntype{L}[1]{>{\raggedright\let\newline\\\arraybackslash}p{#1}} 
\newcolumntype{C}[1]{>{\centering\let\newline\\\arraybackslash}p{#1}} 

\usepackage[utf8]{inputenc}
\usepackage{epstopdf}
\usepackage{tabulary}
\usepackage{booktabs}
\usepackage{float}
\usepackage{xspace}

\begin{document}

\title{Model-Driven Requirements for Humans-on-the-Loop Multi-UAV Missions}

\author{\IEEEauthorblockN{Ankit Agrawal and Jane Cleland-Huang}
\IEEEauthorblockA{Dept. of Computer Science and Engineering} 
University of Notre Dame\\
Notre Dame, IN, USA\\
aagrawa2@nd.edu, JaneHuang@nd.edu
\and
\IEEEauthorblockN{Jan-Philipp Stegh\"ofer }
\IEEEauthorblockA{Dept. of Computer Science and Engineering} 
Chalmers $|$ University of Gothenburg\\
Gothenburg, Sweden\\
jan-philipp.steghofer@cse.gu.se}

\maketitle

\begin{abstract}
The use of semi-autonomous Unmanned Aerial Vehicles (UAVs or drones) to support emergency response scenarios, such as fire surveillance and search-and-rescue, has the potential for huge societal benefits. Onboard sensors and artificial intelligence (AI) allow these UAVs to operate autonomously in the environment. 
However, human intelligence and domain expertise are crucial in planning and guiding UAVs to accomplish the mission. Therefore, humans and multiple UAVs need to collaborate as a team to conduct a time-critical mission successfully. 
We propose a meta-model to describe interactions among the human operators and the autonomous swarm of UAVs. The meta-model also provides a language to describe  the roles of UAVs and humans and the autonomous decisions. We complement the meta-model with a template of requirements elicitation questions to derive models for specific missions. We also identify common scenarios where humans should collaborate with  UAVs to augment the autonomy of the UAVs. 
We introduce the meta-model and the requirements elicitation process with examples drawn from a search-and-rescue mission in which multiple UAVs collaborate with humans to respond to the emergency.  We then apply it to a second scenario in which UAVs support first responders in fighting a structural fire. Our  results  show  that the meta-model and the template of questions support the modeling of  the  human-on-the-loop human interactions for  these  complex  missions, suggesting that  it is a useful  tool  for  modeling  the  human-on-the-loop  interactions for  multi-UAVs missions.
\end{abstract}


\section{Introduction}
The deployment of a swarm of Unmanned-Aerial Vehicles (UAVs) to support human first responders in emergencies such as river search-and-rescue, hazardous material sampling, and fire surveillance has earned significant attention due to advancements in the robotics and Artificial Intelligence (AI) domains \cite{torresen2018review,carpentiero2017swarm}. Advanced AI models can assist UAVs in performing tasks such as creating a 3D heat-map of a building, finding a drowning person in a river, and delivering a medical device,  while robotics autonomy models enable UAVs to automatically plan their actions in a dynamic environment to achieve a task \cite{chung2018survey,hu2020deep}. However, despite these advances, the deployment of such systems remains challenging due to uncertainties in the outcome of the AI models \cite{klas2018uncertainty}, rapid changes in environmental conditions, and emerging requirements for how a swarm of autonomous UAVs can best support first responders during a mission.  

The UAVs of next-generation emergency response systems will be capable of sensing, planning, reasoning, sharing, and acting to accomplish their tasks \cite{nahavandi2017trusted}. These UAVs will not require  humans-in-the-loop to make all key decisions, but rather will make independent decisions with  humans-on-the-loop setting goals and supervising the mission \cite{fischer2017loop}. For example, in a multi-UAV river search-and-rescue mission, the autonomous UAV can detect a drowning person in the river utilizing the on-board AI vision models (\emph{sensing}) and ask another UAV  to schedule delivery of a flotation device to the victim's location (\emph{planning} and \emph{reasoning}). These UAVs collaborate to share (\emph{sharing}) the victim's location and subsequently deliver the flotation device (\emph{acting}). These intelligent UAVs also send the victim's location to emergency responders on the rescue-boat so that they can perform the physical rescue operation. Autonomous systems of such complexity demand humans and intelligent agents to collaborate as a human-agent team~\cite{bellamy2017human,ClelandHuang-iHDI2020}.

A well-known issue in designing a system comprising humans and autonomous agents is to identify how they can collaborate and work together to achieve a common goal~\cite{hancock1998allocating}. The challenges in human multi-agents collaboration include identifying when and how humans should adjust the autonomy levels of agents,  identifying how autonomous agents should adapt and explain their current behavior to maintain humans' trust in them, and finally, identifying different ways to maintain situational awareness among humans and all autonomous agents. In this paper we propose a humans-on-the-loop solution in which humans maintain oversight while intelligent agents are empowered to autonomously make planning and enactment decisions. We first identify common interaction patterns in which humans collaborate with autonomous agents, and then leverage those patterns to  construct a human interaction meta-model. In addition, we define a set of `probing' questions which can be used to elicit, analyze, and ultimately specify requirements for human multi-UAV interactions in specific  emergency response missions. 

This paper makes three primary contributions.  First it motivates the problem of human multi-agent interaction through examples drawn from a concrete mission scenario.  Second, it provides a meta-model to describe human interactions with multiple agents, and finally it presents a set of requirements-related guiding questions for eliciting and then modeling specific instances of these human multi-agent interactions. 

The paper is organized as follows: Section~\ref{sec:motivating_scenarios} presents examples of human multi-agent interactions drawn from the river-rescue scenario and section~\ref{sec:actions} presents an analysis of  these interactions. Section~\ref{sec:meta-model} introduces a human-on-the-loop meta-model for describing human multi-agent interactions. Section  \ref{sec:requirements}  then describe our process for eliciting requirements, mapping them to elements of the meta-model, and then specifying requirements by deriving instances of the meta-model for each identified human multi-agent interaction type. Section \ref{sec:example} discusses an application of our work and finally sections \ref{sec:threats},~\ref{sec:related}, and \ref{sec:conclusion} discuss threats to validity, related work, and draw conclusions.

\section{Human-Multi-UAV Collaborations}
\label{sec:motivating_scenarios}
Several research groups have explored the application of UAVs for specific emergency scenarios such as surveying and assessing damage following an earthquake \cite{XU201422} or volcanic eruption \cite{DEBENI2019250}, investigating maritime spills \cite{DOOLY2016528}, delivering defibrillators \cite{10.1145/2851581.2892288}, and mapping wildfires \cite{Athanasis19}. These applications all involve human operators interacting with UAVs in direct or indirect ways to plan routes, capture video, or to supervise varying degrees of autonomous UAV behavior -- typically through the use of a graphical user interface (GUI).  Researchers have described other forms of interactions \cite{hdi-survey}, including haptic and voice interfaces \cite{funk18,cauchard15}, but these are infrequently used in emergency response applications.  

\subsection{DroneResponse: A Case Environment}
\begin{figure}
    \centering
     {\includegraphics[width=0.98\columnwidth]{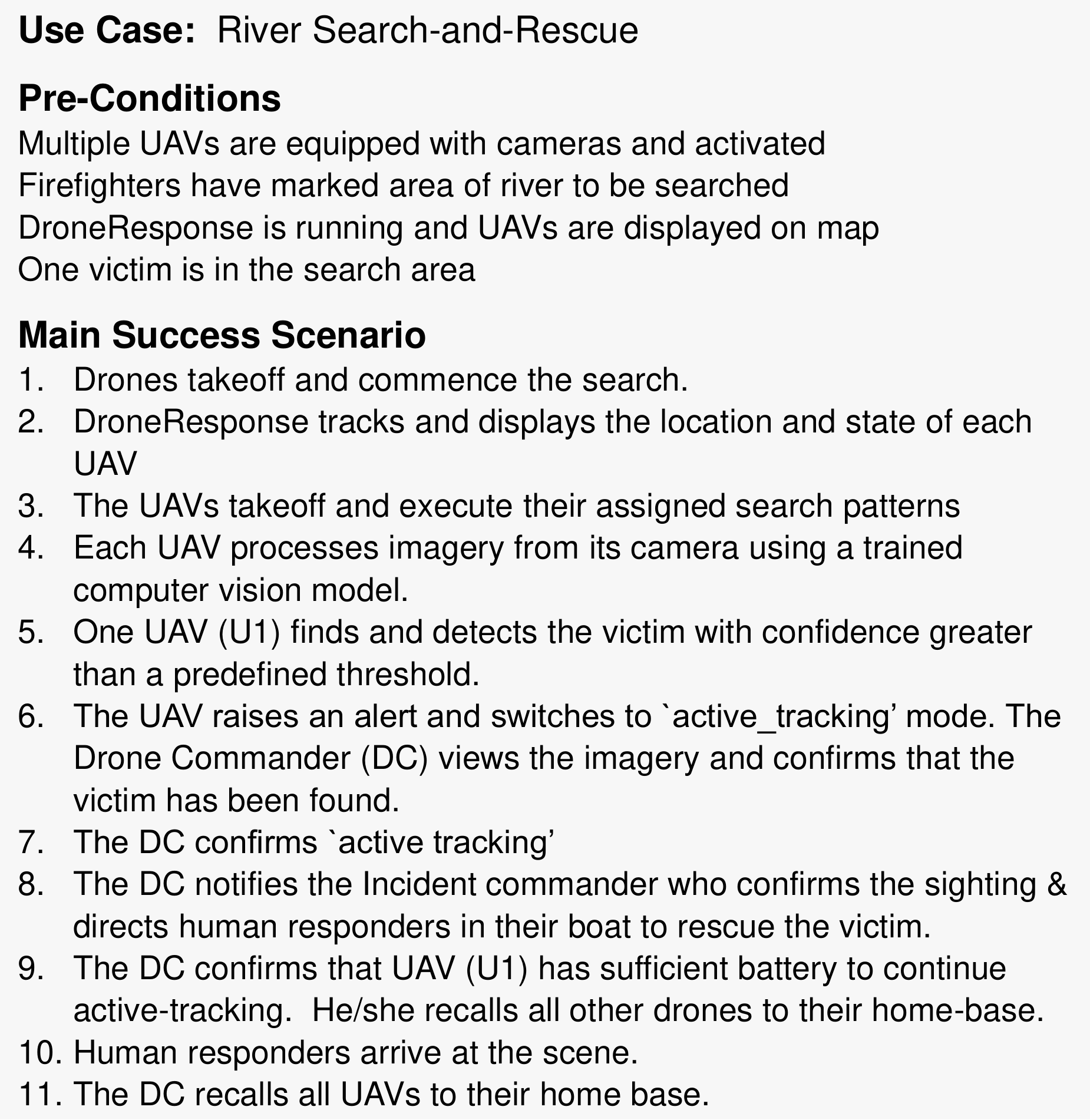}}
     \caption{A partial use case description of the DroneResponse River search-and-rescue scenario.}
     \label{fig:use-case}
    \vspace{10pt}
    \includegraphics[width=0.98\columnwidth]{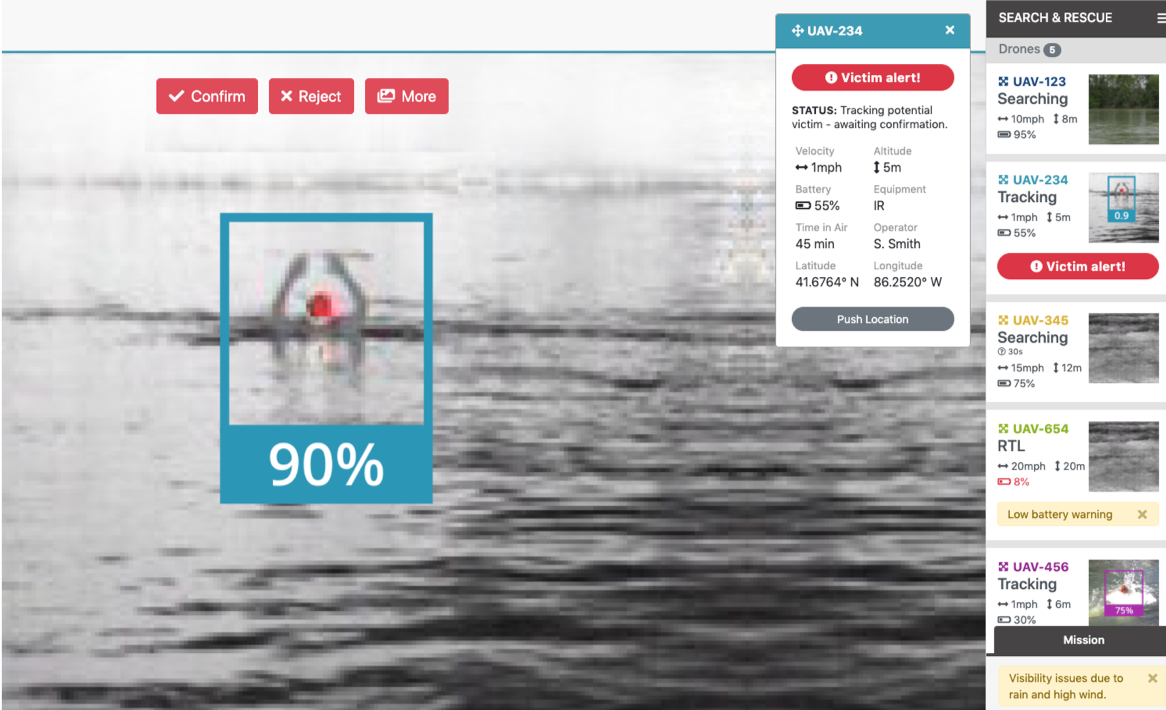}
    \caption{A human-UAV interaction point in which a UAV has detected a candidate victim and requested human confirmation.}
    \label{fig:droneResponse}
    \vspace{-12pt}
\end{figure}

In this paper, we primarily draw examples from our \emph{DroneResponse} system, which we are developing to enable multiple collaborating, semi-autonomous UAVs to support diverse emergency response missions such as fire surveillance, search-and-rescue, and environmental sampling \cite{droneresponse,DBLP:conf/icse/Cleland-HuangVB18,DBLP:conf/euromicro/VierhauserCBKRG18}. Figure~\ref{fig:use-case} depicts a river search-and-rescue use-case in which multiple UAVs are deployed to find a victim on the river and to potentially aid emergency responders in delivering a flotation device.

DroneResponse represents a socio-technical cyber-physical system (CPS) in which multiple humans and multiple semi-autonomous UAVs engage in a shared emergency response mission. UAVs are designed to make autonomous decisions based on their current goals, capabilities, and current knowledge. 
They build and  maintain their knowledge of the mission through directly observing the environment (e.g., through use of their onboard sensors) and through receiving information from other UAVs, central control, and human operators \cite{wooldridge1997agent}. 
UAVs then work to achieve their goals through enacting a series of tasks \cite{pokahr2005jadex}. 

Humans interact with UAVs through various GUIs to create and view mission plans, monitor mission progress, assign permissions to UAVs, provide interactive guidance, and to maintain situational awareness. Bidirectional communication is crucial for enabling both humans and UAVs to complement each other's capabilities during the mission.  An example of human-UAV collaboration is depicted in Figure~\ref{fig:droneResponse}, which shows a UI developed for the DroneResponse system.  In this example, the UAV has detected a candidate victim in the water, autonomously started tracking the victim, while simultaneously  requesting confirmation from the human incident commander that the detected object is actually the victim. 
\subsection{Human-UAV Interactions}
\label{sec:human-uav-interactions}
DroneResponse is being developed in close collaboration with emergency responders through engagement in a series of brainstorming activities, interviews, participatory design sessions, and early field-tests~\cite{droneresponse,DBLP:conf/icse/Cleland-HuangVB18,DBLP:conf/re/Cleland-HuangV18}.  
The following concrete examples of human-UAV interactions, taken from the river search-and-rescue example, were identified as part of this collaborative design process. We use these examples throughout the remainder of the paper to motivate and contextualize our modeling activities. \newline \vspace{-8pt}

\noindent{\bf Scenario S1 -- Planning a rescue strategy:~}
When a UAV identifies a potential victim in the river, the victim's coordinates are sent to the mobile rescue unit. However, the UAV must also decide whether to request delivery of a flotation device by a suitably equipped UAV or whether it is sufficient  to simply continue streaming imagery of the victim until human rescuers arrive.  The UAV makes this decision by estimating the arrival time of the rescue boat versus the time to deliver a flotation device. However, humans can contribute additional information to the decision -- for example, by modifying the expected arrival time of the rescue boat, or by inspecting the streamed imagery and determining whether the victim would be able to receive the flotation device if it were delivered (e.g., the victim is conscious and not obscured by overhead branches) and is in need of the device (e.g., not having a safe waiting position on a rock or tree branch).  This is an example of a \emph{bidirectional exchange of knowledge} between multiple humans and multiple UAVs, where the first UAV shares the victim's coordinates and streams imagery, humans on the boat estimate their ETA and if necessary update the UAV's situational awareness, the incident commander decides whether a flotation device could be used effectively if delivered on time, and if needed, a second UAV performs the delivery. The scenario illustrates many aspects of human-agent collaboration including \emph{knowledge sharing} and \emph{human intervention}.   \newline \vspace{-8pt}

\noindent{\bf Scenario S2 -- Sharing environmental information:~} In river search-and-rescue missions, victims tend to get trapped in `strainers' (i.e., obstruction points) or tangled in tree roots on outer banks. These areas require closer inspection. While UAVs have onboard vision and will attempt to identify `hotspots', human responders can directly provide this information to multiple UAVs based on their observation of the scene. This enables UAVs to collaboratively adapt their flight plan so that they prioritize specific search areas, or adjust their flight patterns to reduce speed or fly at lower altitudes in order to render higher-resolution images of priority search areas. This interaction scenario is similar to the previous one, except that it is primarily uni-directional with information passed from humans to UAVs.  \newline \vspace{-8pt}

\noindent{\bf Scenario S3 -- Victim confirmation:~} The UAV's AI model uses its onboard computer vision to detect potential victims. When the confidence level surpasses a given threshold, the UAV will autonomously switch to tracking mode and broadcast this information to all other UAVs. If the UAV autonomy level is low, it requests human confirmation of the victim sighting before it starts tracking. Human feedback is sent to the UAV and propagated across all other UAVs.  In this scenario the \emph{UAV elicits help from the human} and the human responds by confirming or refuting the UAV's belief that it has sighted a victim or by suggesting additional actions. For example, if the detected object is partially obscured, the human might ask the UAV to collect  additional imagery from multiple altitudes and angles. \newline \vspace{-8pt}

\noindent{\bf  Scenario S4 -- Support for UAV coordination:~} 
In an extension to the previous scenario, multiple UAVs might simultaneously detect a victim.  They must then use onboard computer vision and their own estimated coordinates of the detected object to determine whether they have detected the same object and to plan a coordinated response. However, this determination may be more complicated in poor visibility environments with weak satellite signals and low geolocation accuracy (e.g., in canyons). Human responders may need to intervene in the UAV's planning process by helping determine whether the sighted objects are valid and unique, and if necessary selecting the most appropriate UAV for the tracking task. This is an example in which the human \emph{intervenes in the UAV's autonomy} and potentially provides \emph{direct commands}, assigning a specific UAV to the task.  \newline \vspace{-8pt}

\noindent{\bf  Scenario S5 -- Prohibiting normal behavior:~} Most UAVs come with built-in safety features so that they autonomously land-in-place or return to launch (RTL) when their battery becomes low or a malfunction is detected. In the case of a low battery, the DroneResponse system initially raises a low-battery alert in the UI, and eventually initiates the RTL command. A human responder might modify the UAV's permissions and \emph{prohibit the UAV from transitioning to RTL} if the UAV is conducting a critical task. An example, that arose from discussions with the Navy, was the use of floating drones for man-overboard scenarios. If a UAV found a victim, and no other UAV or human rescue unit were in the vicinity, the RTL feature would be deactivated automatically. This meant that when batteries lost power, the UAV would land in the water and serve as a search beacon.   However, for many reasons, a human might wish to override the default deactivation of the RTL, thereby reactivating the UAV's RTL autonomy. 

These motivating examples provide the foundation for our discussion of human-on-the-loop collaboration patterns. 


\section{Analysis of Collaboration Actions}
\label{sec:actions}

\begin{figure}[!t]
    \centering
    \includegraphics[width=\columnwidth]{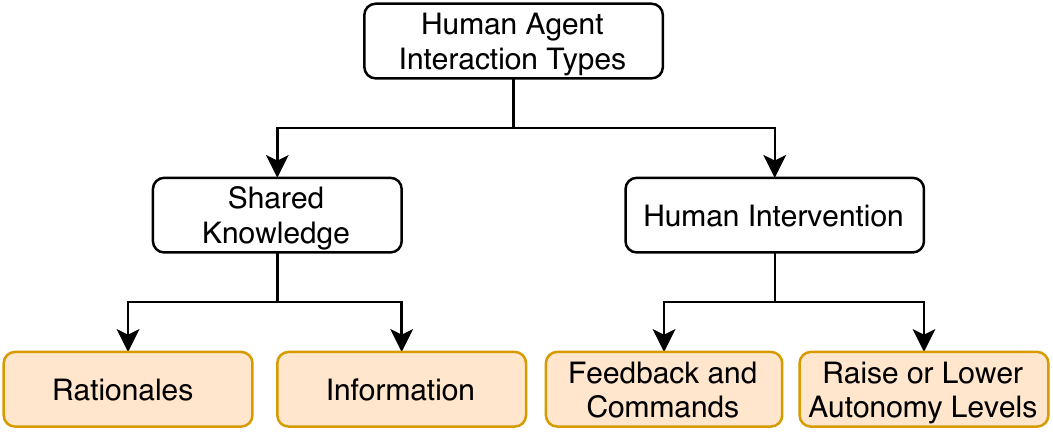}
    \caption{Humans and agents collaborate through shared knowledge and through human interventions in the agents' autonomy.}
    \label{fig:HumanCollaborationPoints}
\end{figure}

Agents within a human-on-the-loop (HotL) system are empowered to execute tasks independently with humans serving in a purely supervisory role~\cite{scharre2015introduction}. However, as our previous examples have shown, humans and agents continually share information in order to maintain bidirectional situational awareness and to work collaboratively towards achieving  mission goals. Agents report on their status (e.g.,  remaining battery levels, GPS coordinates, and altitude), and they  explain their current plans, actions, and autonomous decisions whenever requested by humans.  Humans can directly intervene in the agents' behavior by providing additional information about the environment, and agents can then leverage this information to make more informed decisions. Humans also respond to direct requests for feedback -- for example, to confirm a victim sighting as previously discussed. They can also provide direct commands (e.g., RTL or stop tracking), or can explicitly modify an agent's permissions in order to enhance or constrain the agent's autonomous behavior. These types of interactions are depicted in  Figure~\ref{fig:HumanCollaborationPoints}. 

\subsection{Situational Awareness}
Situational Awareness (SA) is the ability of the user to perceive the environment (Level-1 SA), to understand the reasoning behind the current state of the environment (Level-2 SA), and finally, to project how the situation could evolve in the future (Level-3 SA)~\cite{ensdley_SA}. 
Humans acquire knowledge of the situation from diverse sources such as their physical interactions with the agents (e.g., visual observations and sounds), observations of the current weather, radio communication with on-scene first responders, and finally through information shared through the systems' GUI. Humans combine knowledge from all of these sources to create a mental model of the current status of the mission.  At the same time, autonomous agents, such as UAVs, develop their own situational awareness using their onboard sensors and through collating information shared by other autonomous agents and by humans. Both humans and autonomous agents then use their shared knowledge of the environment to formulate and enact plans to collaboratively achieve their mission goals. 

In a HotL environment, agents make many autonomous decisions; however, in order for humans to supervise the mission and to maintain full situational awareness, the agents must explain their behavior when requested by a human. The explanation should include key \textbf{information} (i.e., the agent's situational awareness at the time the decision was made), the autonomous decision (e.g., switch modes, change altitude etc), and a human understandable rationale for the decision. Providing {\bf rationales for all decisions and subsequent behavior} is therefore critical in order for humans to achieve situational awareness. If the human were to disagree with the decision and the logic of the supporting rationale, then they could monitor the agents more closely, temporarily lower their autonomy levels, or make longer-term adjustments (e.g., retraining a computer vision model) for future missions.


\subsection{Human Intervention}

At times, humans may need to intervene in the autonomy of an agent in order to influence and improve the outcome of the joint mission. They can do so in several different ways. 
Previous studies\cite{loftin2016learning, thomaz2006reinforcement} demonstrate that a {\bf feedback loop} can help agents to improve their future performance by fine-tuning algorithmic parameters that drive the agent's autonomy. For example, feedback on a candidate victim detected by the computer vision model, could be used to retrain the model or refine its configuration parameters, thereby potentially reducing false positives or false negatives. In addition, users can initiate commands to  immediately enact changes in the behavior of the UAV. For example, a human could directly command a UAV to fly to a specific waypoint to checkout a report received on social media.  

Finally, the human may choose to {\bf raise or lower autonomy levels} of the agent. Autonomy levels, defined as the extent of an agent's independence while acting autonomously in the environment, can be expressed through role assignments or through specific permissions within a role. For example, a UAV that is permitted  to track a victim without first obtaining  human confirmation has a higher autonomy level than  one which needs explicit human confirmation before tracking. Humans tend to establish autonomy levels based on their trust in the agent's capabilities. For example, a UAV exhibiting high degrees of accuracy in the way it classifies an object increases human trust, and as a result, the human might grant the UAV additional permissions.  On the other hand, the human operator might revoke permissions, thereby lowing autonomy levels, if the UAV were operating in weather conditions for which the computer vision model had not been appropriately trained and for which accuracy was expected to be lower than normal.

\section{Meta-Model for Human-UAV Interactions}
\label{sec:meta-model}
We constructed a meta-model to define the vocabulary of the domain of human multi-agent interactions. The meta-model includes domain-specific  concepts and establishes rules for how those types of concepts are associated with one another. This allows us to express specific instances of human multi-agent interaction in conceptual models and reuse the concepts we identified to express how humans and multiple agents will interact with each other in specific scenarios.

\begin{table}[h!]
	\centering
	\caption{Additional Use-Cases from which human multi-UAV interaction patterns were identified and analyzed}
	\label{tab:usecases}
        \addtolength{\tabcolsep}{-4.5pt}
        \small
        \begin{tabular}{L{.8cm}L{4cm}L{4cm}@{}}
        \hline
          {\bf ID} & {\bf Use Cases} & {\bf Engaged Stakeholders}\\ \hline
             UC1  & River Search \& Rescue &South Bend Firefighters\\ 
             UC2  & Defibrillator Delivery &DeLive, Cardiac Science\\ 
             UC3  & Traffic Accident surveillance&South Bend Firefighters\\ 
             UC4  & Water Sampling  &Environmental Scientists\\ 
        UC5&Man overboard& US Navy\\ 
        \hline
        \end{tabular}
         
      \end{table}

\begin{figure}
    \centering
    \includegraphics[width=\columnwidth]{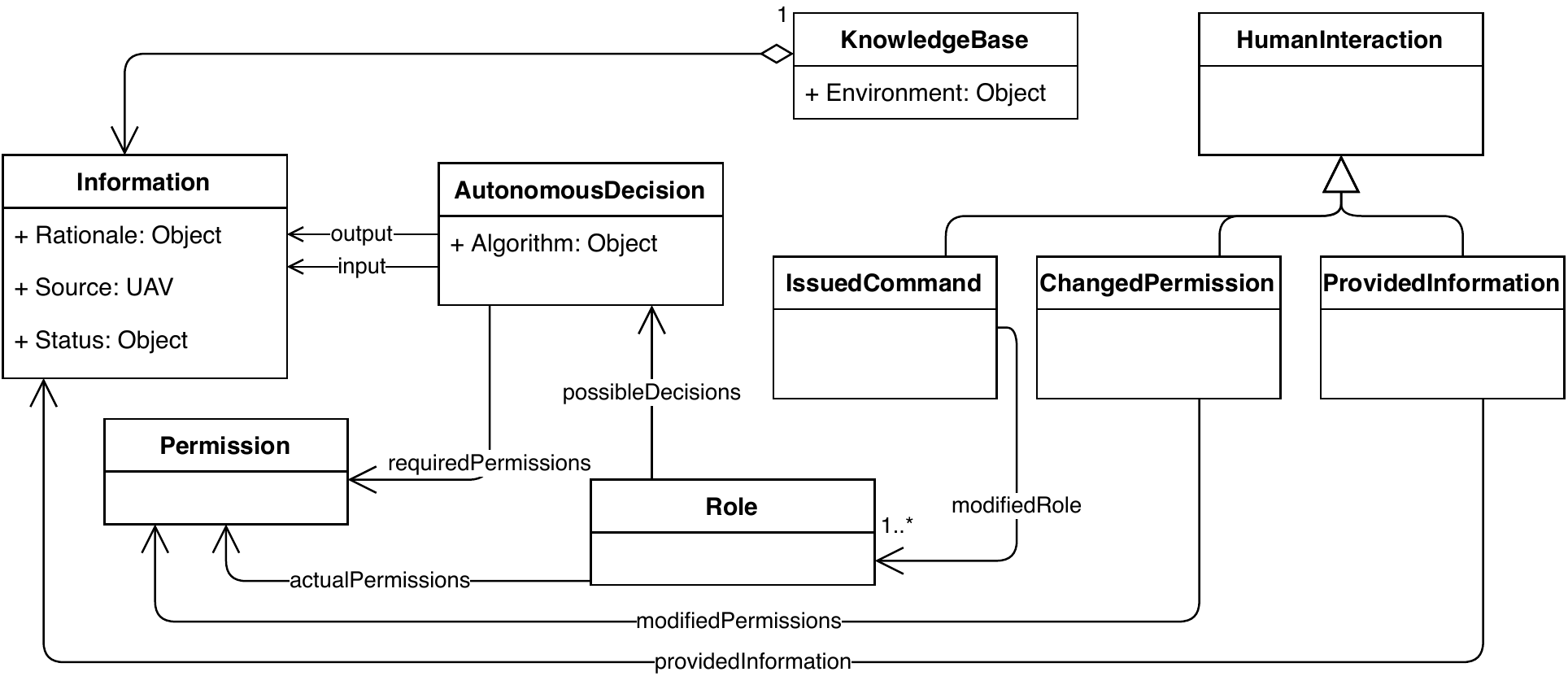}
    \caption{Meta-model for human-on-the-loop interaction in a multi-agent mission.}
    \label{fig:human-on-the-loop-meta-model}
\end{figure}

The elements of the meta-model were (cf.~Fig.~\ref{fig:human-on-the-loop-meta-model}) derived from our analysis of human multi-UAV interactions in the river-rescue scenarios and also from additional scenarios summarized in Table \ref{tab:usecases}. The meta-model depicts frequently occurring concept types and their associations, and was designed iteratively through multiple refinements in which we recursively validate the model against the specific scenarios described in Section \ref{sec:motivating_scenarios}.  Our meta-model includes the following elements:

A \texttt{Role} defines the complex behaviors that agents perform autonomously. Complex behaviors of a UAV include takeoff, search, track, deliver, and RTL. 

An \texttt{AutonomousDecision} entity uses algorithms that leverage \textit{Information} in the \texttt{KnowledgeBase} to make decisions. The complex behaviour of a \texttt{Role} is defined through one or several such decisions.
For example, there are many cases in which a single agent must serve as a \textit{leader}, responsible for coordinating behavior of its \textit{follower}s. During a leader election, an \texttt{AutonomousDecision} entity could select a new leader from the set of followers, thereby enabling the system to switch leaders without the need for human intervention. Upon making a decision, an \texttt{AutonomousDecision} entity generates output \texttt{Information} including a rationale for its decision, which could later be used to generate a human-readable explanation.

Entities of type \texttt{Permission} are used by \texttt{AutonomousDecision}s to decide if the agents are allowed to make a specific decision. For example, an \texttt{AutonomousDecision} entity checks whether the human responders have allowed the system to automatically select a replacement if needed during a victim tracking activity. \texttt{Role}s are associated with a set of permissions defining the allowed behaviors of the agent which can be modified at run-time.

A \texttt{KnowledgeBase} entity contains current environmental information as well as information about the state of a single agent or multiple agents. An \texttt{AutonomousDecision} entity uses the \texttt{Information} stored in the \texttt{KnowledgeBase} in decision making. A human can use the information in the \texttt{KnowledgeBase} entity to gain situational awareness of the mission.

Entities of type \texttt{HumanInteraction} allow humans to intervene in the autonomy of the agents or to share their explicit knowledge of the environment.
The three entity types \texttt{ProvidedInformation}, \texttt{ChangedPermission}, and \texttt{IssuedCommand} provide different ways for humans to interact with the system. The \texttt{ProvidedInformation} entity adds \texttt{Information} to the \texttt{KnowledgeBase} of the system to maintain the consistent knowledge among multiple agents. Humans can use interventions of type \texttt{ChangedPermission} to raise or lower the autonomy of an agent, or agents, based on their trust in the ability of  the agents to make correct decisions within the current environment.  Finally, an \texttt{IssuedCommand} entity allows humans to gain control over the autonomous behavior of the agents. For example, if a UAV loses communication with other UAVs in the mission and fails to deliver the flotation device when it is needed, a human can send a direct command that sets the current \texttt{Role} of the UAV to \textit{deliver flotation device}.

It is noteworthy that neither humans nor agents are represented explicitly in our meta-model. The underlying implicit assumption is that roles are assigned to agents according to the capabilities of each UAV, that UAVs can assume new roles according to the state of the environment, constrained by permissions associated with their capabilities. Furthermore, humans and agents have access to one or several instances  of the distributed \texttt{KnowledgeBase} which stores information acquired from the environment, multiple UAVs, and from humans. The reason for leaving these aspects implicit are that the domain of our model is human multi-UAV interaction and it is not relevant to the meta-model to specify which concrete UAV has assumed each specific role.

\section{Requirements Modeling}
\label{sec:requirements}
Human multi-agent  interactions in the domain of emergency missions are impacted by factors such as uncertainty of the agents' knowledge, the degree of human trust in the agent's ability to reason over its knowledge and behavior correctly, and the criticality of the task at hand.  Autonomy levels and human interactions should therefore not be applied at the same level for all tasks, in all contexts, and across all phases of the mission, but instead need to be customized according to actions, context, phase, and even human preferences. This introduces the need for a systematic requirements elicitation process to explore the knowledge needs of humans and agents, and  identify points at which humans can interact with the agents' autonomous behavior.  

To support the elicitation, analysis, and specification of human multi-agent interactions, we developed a  set of probing questions~\cite{anish2016probing,miller2009quest_probingQuestions}.  These questions can be used to elicit requirements  for each human multi-agent interaction point from system stakeholders. Probing questions are not necessarily easy to answer especially as human multi-agent interactions represent an emergent area of study with unknown unknowns~\cite{DBLP:conf/re/SutcliffeS13}. Answering the questions therefore requires a rigorous and systematic requirements engineering elicitation and analysis process that includes brainstorming, interviews, immersive prototyping, and even field-studies in order to fully discover the requirements~\cite{DBLP:journals/ijmms/Robertson01, DBLP:conf/re/Sutcliffe01}. 

We structure our probing questions around the four types of human multi-agent interactions defined in  Figure~\ref{fig:HumanCollaborationPoints}.  These include (1) information sharing, (2) direct feedback and commands, (3) raising or lowering of autonomy levels, and (4) providing behavior rationales and explanations. We map each question to the entities of the meta-model, and then use the answers  to specify the requirements for each interaction point as a conceptual model. In each case, the first question is designed to identify specific interaction points, while all subsequent questions are used to explore the details of each interaction. 

\subsection{Sharing Information}
At the most basic level, humans and agents must share information with each other in order to create a common understanding of the state of the mission and its environment.  We therefore start by posing two key questions concerning the exchange of information. 

\begin{enumerate}[leftmargin=.75cm]
\item [{\footnotesize PQ1}:] {\bf What information} do agents or humans need to know about the state of the mission and the environment in which they operate  individually or collaboratively?
[\texttt{\small Knowledge, Role, AutonomousDecisions}]
\item [{\footnotesize PQ2}:] When and how will these agents or humans share or acquire information?
[\texttt{\small Knowledge, Information, Role}]
\end{enumerate}

By default, the system must be designed such that information is shared freely across humans and agents.  For example, agents acquire knowledge about the environment and the state of the mission through their sensors (e.g., victim detected or wind velocity 20 mph) and through decisions they make (e.g., UAV-1 is tracking a detected victim). They share this information with other active agents and with humans on the ground. However, above and beyond this general exchange of information, we must explore additional explicit  interaction points between humans and agents in order to understand the system's requirements.  

\subsection{Feedback and Commands}
All five of the scenarios in Section~\ref{sec:human-uav-interactions} introduce the possibility of a human offering feedback or even direct commands.  To elicit a more complete list of interaction points, we ask the following question:

\begin{enumerate}[leftmargin=.75cm]
\item [{\footnotesize PQ3}:] When should a {\bf human intervene} by providing direct feedback or commands to multiple agents? 
[\texttt{\small IssuedCommand, AutonomousDecision, ProvidedInformation}]
\end{enumerate}
We then ask additional probing questions to explore each of the identified intervention points:
\begin{enumerate}[leftmargin=.75cm]
\item [{\footnotesize PQ4}:] What {\bf triggers} the feedback or command? ~(e.g., solicited by UAV, triggered by a specific event, or offered by the human operator based on his/her general awareness) 
[\texttt{AutonomousDecision, Information}]

\item [{\footnotesize PQ5}:] What {\bf information} should be provided in the feedback or command? ~(e.g., knowledge of the scene, permission to perform a specific task, a hint) 
[\texttt{\small Information}]

\item [{\footnotesize PQ6}:] How should the agent {\bf respond to the feedback}? ~(e.g., update its situational awareness, obey the command regardless of its current environmental knowledge) 
[\texttt{\small Role, AutonomousDecisions}]
\end{enumerate}

\subsection{Providing behavioral rationales}
Scenarios S4 and S5 provided clear examples in which a UAV needed to explain its behavior. To identify other such interaction points we pose the following question: 
\begin{enumerate}[leftmargin=.8cm]
\item [{\footnotesize PQ7}:] In what concrete situations would humans require agents to explain themselves? 
[\texttt{\small AutonomousDecision}]
\end{enumerate}
The following questions are then posed for each situation in which the agent is expected to explain its behavior.\vspace{3pt}
\begin{enumerate}[leftmargin=.9cm]
\item [{\footnotesize PQ8}:~] Why does the agent need to {\bf explain itself} at this collaboration point? ~(e.g., unexpected behavior) 
[\texttt{\small Role}]
\item [{\footnotesize PQ9}:] What {\bf information} needs to be included in the explanation?  (e.g., current task, goals, actions, rationales) 
[\texttt{\small Information}]

\item [{\footnotesize PQ10}:] Under what circumstances might the human choose to override the agent's decision based on its explanation? If so, what would those overrides look like? (e.g., feedback/command, or lowering of autonomy levels.)
[\texttt{\small HumanInteraction, ProvidedInformation, IssuedCommand, ChangedPermission}]

\end{enumerate}

\subsection{Raising or Lowering of Autonomy Levels}
Scenarios S4 and S5 also provide examples where a human operator may wish to raise or lower autonomy levels. To identify such intervention points we pose the following question: 
\begin{enumerate}[leftmargin=.9cm]

\item [{\footnotesize PQ11}:] When and where do the agents exhibit {\bf autonomous decision-making behavior?} [\texttt{\small Role, Autonomous Decision} ]
\end{enumerate}
Each identified intervention point is then explored through the following questions:
\begin{enumerate}[leftmargin=.9cm]
\item [{\footnotesize PQ12}:] {\bf What information} do the agents need in order to exhibit the autonomous behavior?  [\texttt{\small Information}]

\item [{\footnotesize PQ13}:] Under {\bf normal operating conditions}, what decisions should the agent be able to make autonomously? 
[\texttt{\small AutonomousDecision}]

\item [{\footnotesize PQ14}:] What {\bf constraints} on the agent's autonomy are introduced by issues related to safety, ethics, regulatory requirements, or human trust? (e.g., FAA Part 107 regulations prohibit night-time flight without an explicit waiver) 
[\texttt{\small Permission}]

\item [{\footnotesize PQ15}:] How is the {\bf autonomy suppressed or increased} at this interaction point? (e.g., modifying the confidence threshold for automatically tracking a potential victim, disabling/enabling the ability to track without permission, disabling/enabling the ability for a UAV to determine its ideal altitude and velocity during a search -- or altering the range of allowed values.) 
[\texttt{\small Role, ChangedPermission}]

\item [{\footnotesize PQ16}:] Are there circumstances in which the human needs to make run-time decisions about suppressing or raising autonomy (i.e., human interaction is required) vs. clearly defined rules by which the autonomy levels can be automatically raised and lowered? 
[\texttt{\small Permission, ChangedPermission}]

\item [{\footnotesize PQ17}:] When autonomy is suppressed or increased what {\bf extra support structures} would be needed, if any, for the emergency responders? (e.g., the operator manually pilots multiple UAVs and additional 360$^o$ views are needed).
[\texttt{\small Role}]
\end{enumerate}

\subsection{Constructing Requirements Models}
For each identified human multi-agent interaction point we specify requirements for the interaction by constructing a conceptual model showing named instances of each entity and the relationships between them.
We use the tags assigned to each probing question to identify entities to include in the diagram. We also use the relationships depicted in the meta-model to guide the addition of appropriate relations among the entities. 

\begin{figure}[!t]
    \centering
    \includegraphics[width=\linewidth]{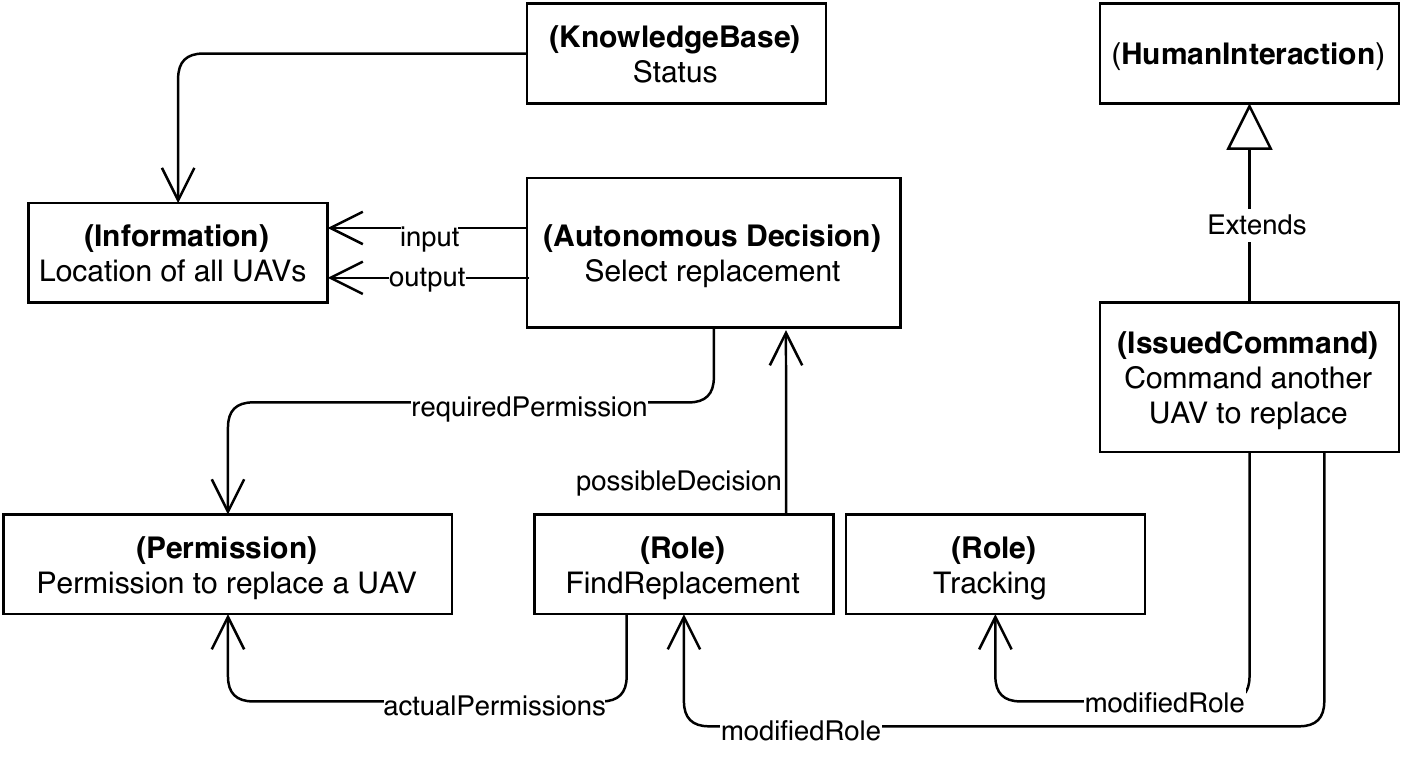}
    \caption{The conceptual model of human intervention in the automatic selection of a UAV to replace another UAV. A human operator issues commands to modify the role of a UAV, thereby overriding autonomous decision of the system. }
    
    \label{fig:M1_instance}

    \includegraphics[width=\linewidth]{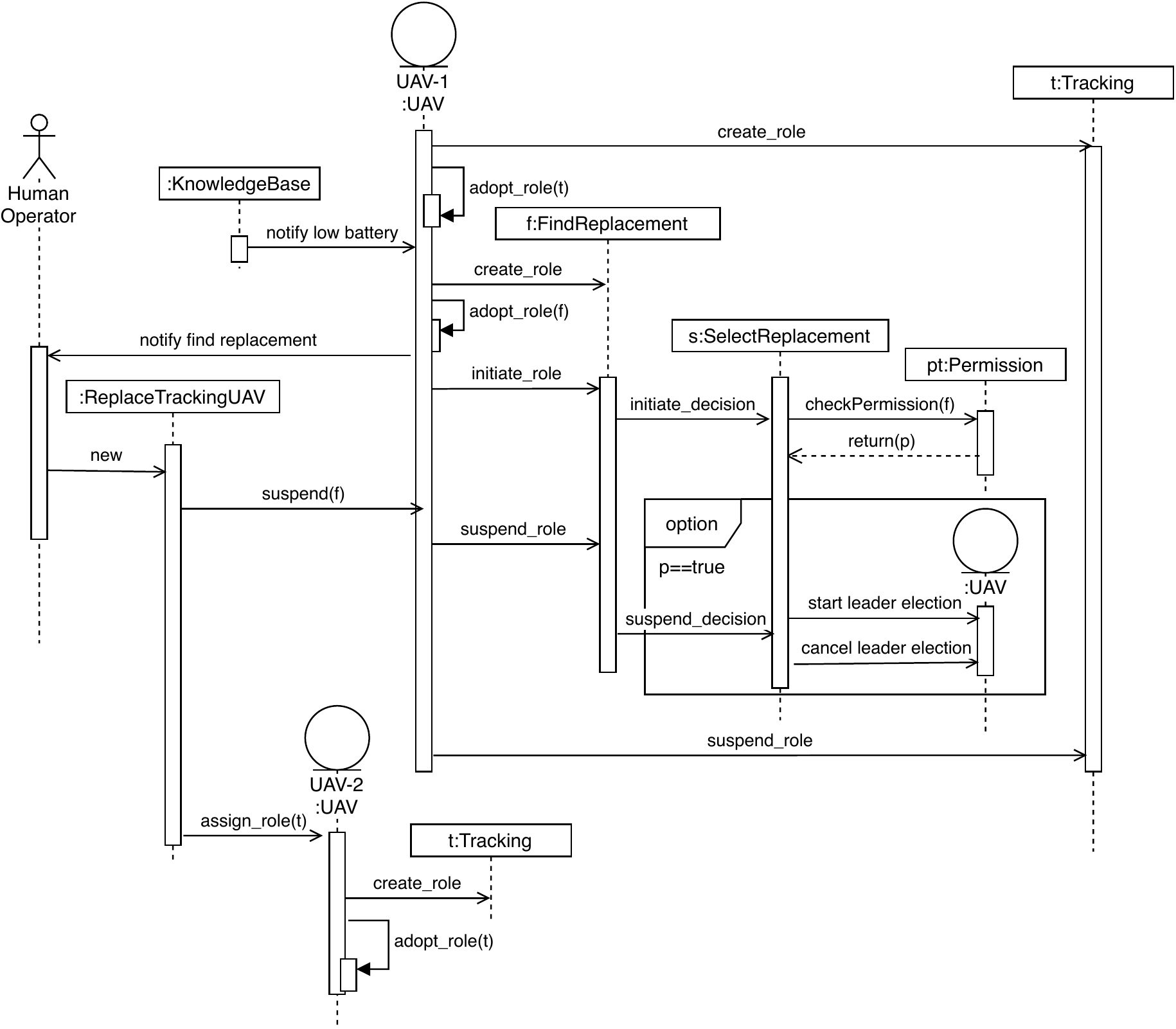}
    \caption{The sequence of events when the human operator intervenes to override an autonomous decision. UAV-1 begins the process of selecting a replacement. However, the human operator overrides the decision by assigning a tracking role to UAV-2 and at the same time suspending UAV-1's search for a replacement.}
    \label{fig:sequence_replace}

\end{figure}

We illustrate the construction of the concept models following the template of probing questions with an example from the river search-and-rescue scenario. The constructed model is shown in Fig \ref{fig:M1_instance}. Probing question \textit{\small PQ11} identifies an example of autonomous behavior that occurs when the battery level of a UAV performing a critical task (e.g., tracking) falls below a predefined level. By default, the UAV will automatically RTL; however, it first requests a replacement from other UAVs in the mission. Therefore, \textit{\small PQ11} identifies the \textit{FindReplacement} role of a UAV. The other UAVs in the mission must autonomously and collaboratively \textit{select a replacement} for the tracking task. \textit{\small PQ12} identifies the required information (\textit{location of all UAVs}), while \textit{\small PQ14} and \textit{\small PQ15} identify the \textit{permission levels} a UAV needs in order to serve as a replacement for the tracking task. \textit{\small PQ3} also  reveals that  human responders reserve the right to override the choice of UAV for any reason, identifying a new command to \textit{replace UAV}. 
Consequently, \textit{\small PQ6} clarifies that the targeted UAV must perform  the \textit{tracking} task after receiving the replacement command from a human responder. In this way, the probing questions help to identify  entities from the meta-model that are required to model this specific human interaction.  We then leverage the relationships between entity types defined in the meta-model to construct a conceptual model of the human multi-UAV interaction in the river search-and-rescue  scenario as shown in Figure \ref{fig:M1_instance}. Finally, we leverage the conceptual model to explore and specify the sequence of events for the human interactions. This entire scenario is depicted in the Sequence Diagram of Figure~\ref{fig:sequence_replace}. 
\section{Application: Structural Fire Support}
\label{sec:example}
As previously described, we constructed our meta-model based on examples from river-rescue and other scenarios shown in Table \ref{tab:usecases}. In this section we briefly illustrate that the proposed meta-model and the probing questions can be used to specify requirements for other human multi-agent use-cases such as structural fire support. We collected an initial set of requirements for this scenario during a series of brainstorming sessions with the South Bend firefighters in the spring of 2019. The firefighters had already used manually-flown UAVs to support their firefighting efforts; and our brainstorming session focused on how they would extend their current use-case to leverage semi-autonomous UAVs as part of our DroneResponse system. 

\begin{figure}[!ht]
    \centering
    \includegraphics[width=\columnwidth]{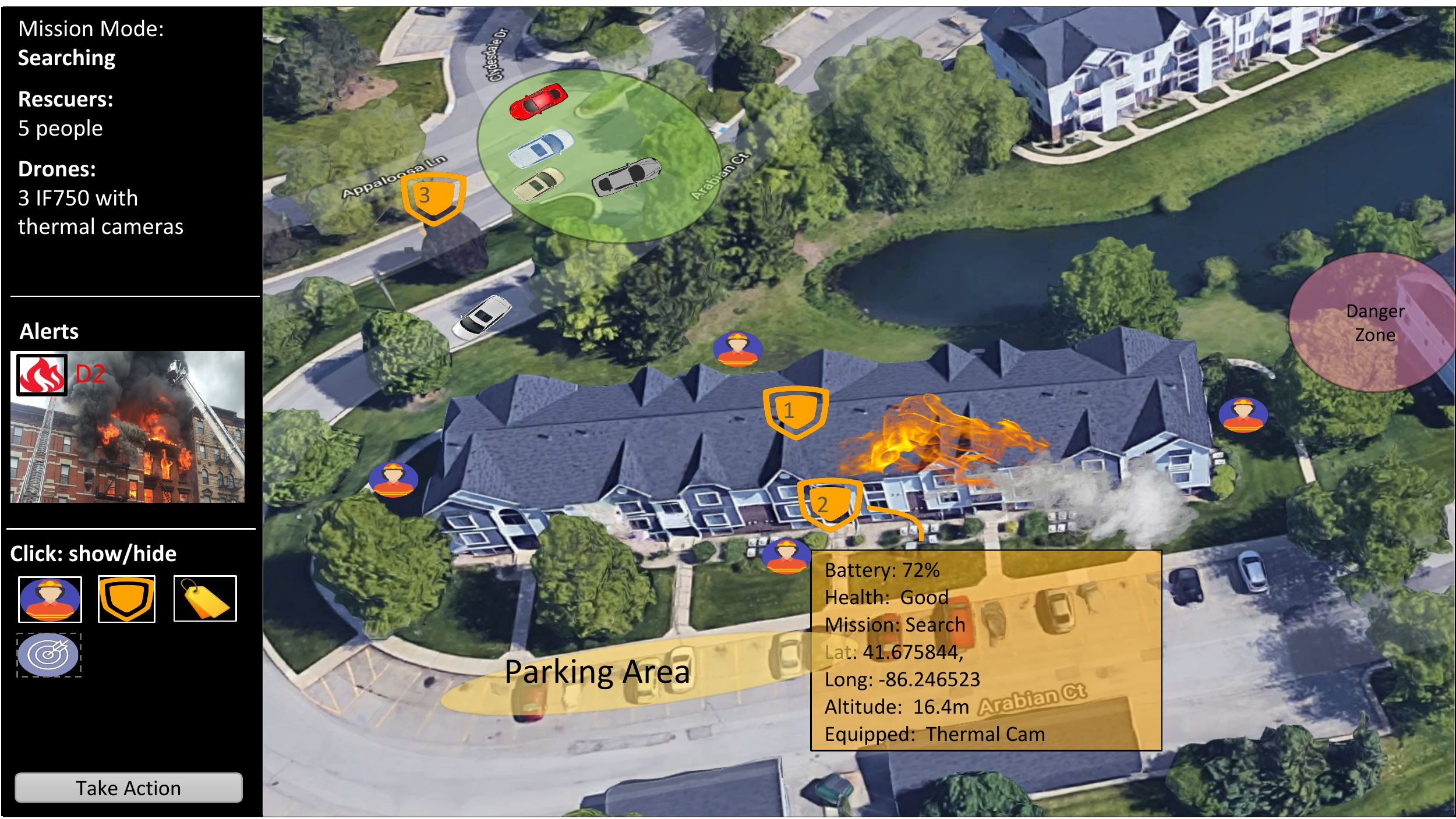}
    \caption{A visionary prototype that we used during brainstorming meetings with Fire Fighters to trigger ideas about the use of DroneResponse in fighting structural fires.}
    \label{fig:fire_mokeup}
\end{figure}

For the purposes of this paper, we leverage the feedback we acquired during the previous brainstorming sessions to retroactively answer the probing questions and to provide an additional example of modeling human interaction requirements.  
Figure \ref{fig:fire_mokeup} shows a visionary mockup used in our original brainstorming session to encourage discussion about the use of UAVs in firefighting. The firefighters identified two primary use cases.  First, they wanted to use UAVs to create thermal maps of the building -- focusing especially on detecting hotspots on roofs as many firefighters have been injured when a roof has collapsed without warning due to an undetected internal fire. They even suggested that UAVs could mark hotspots with lasers. Second, they proposed using UAVs to search for victims through windows and smoke using thermal cameras. 


To demonstrate that our meta-model can be applied to this very different scenario we focus on a specific fire-fighting scenario in which multiple UAVs work collaboratively to create a 3D model of the building. At the start of the mission, the UAVs collaboratively create a plan for surveying the building. For example, depending upon the size and layout of the building, weather conditions, and the number of available UAVs, they could work independently on different parts (sides, roof) of the building, they could prioritize specific areas, fly around the building in either direction, or even work together on a single section at distinct altitudes.  In the scenario that we model, the UAVs devise a specific mapping plan; however, firefighters observe smoke coming from a different area of the building, update the knowledge base, and this leads to the UAVs redesigning their strategy.  In this example, the firefighters do not issue a direct command, but instead provide additional information and allow the UAVs to autonomously adapt their plans. In this example, one of the UAVs assumes a new role of using thermal imagery to search for victims through windows in the area at which smoke has been detected. 

\begin{figure}
    \centering
    \includegraphics[width=\linewidth]{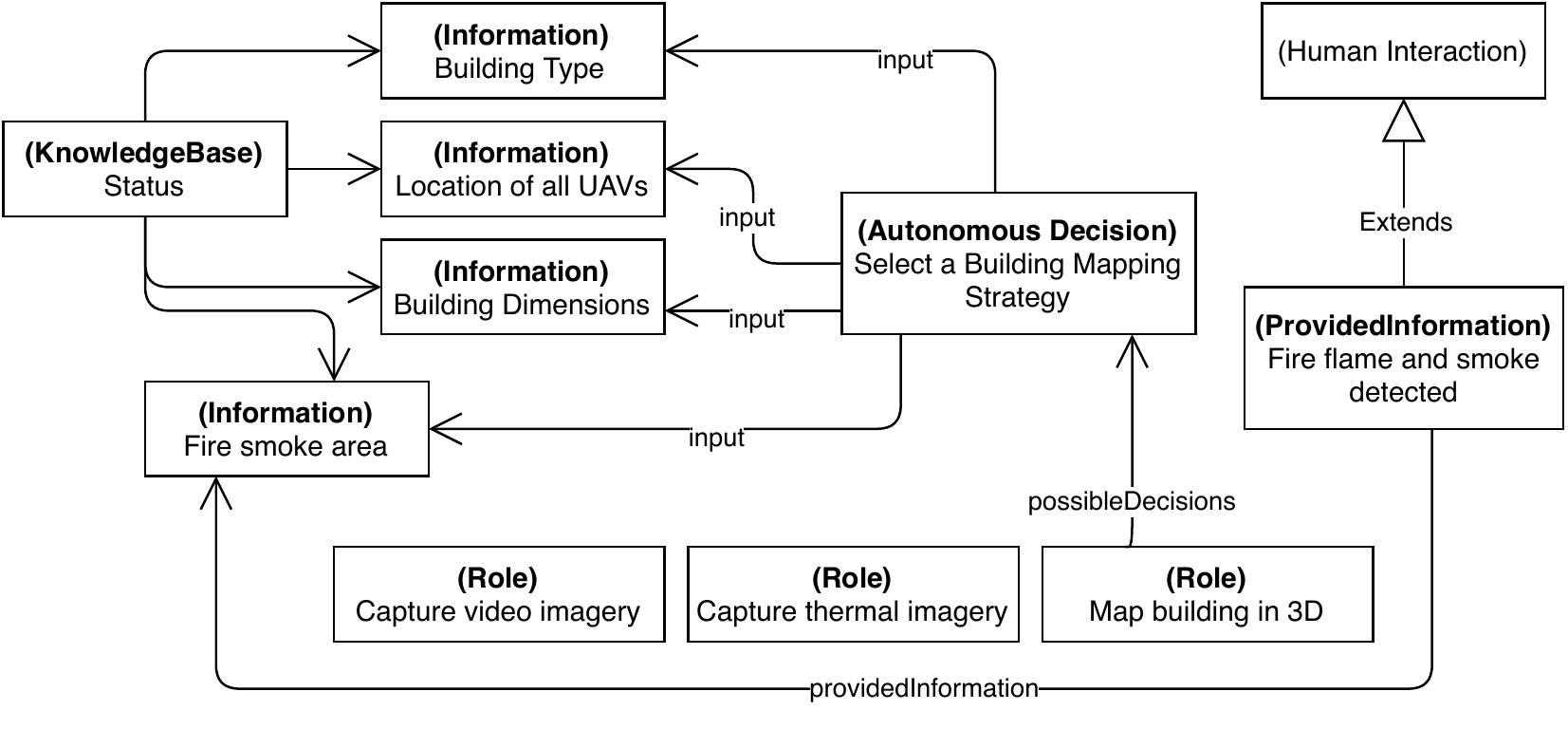}
    \caption*{a: Conceptual model showing entities involved in the human multi-agent interaction.}
    
    \includegraphics[width=\linewidth]{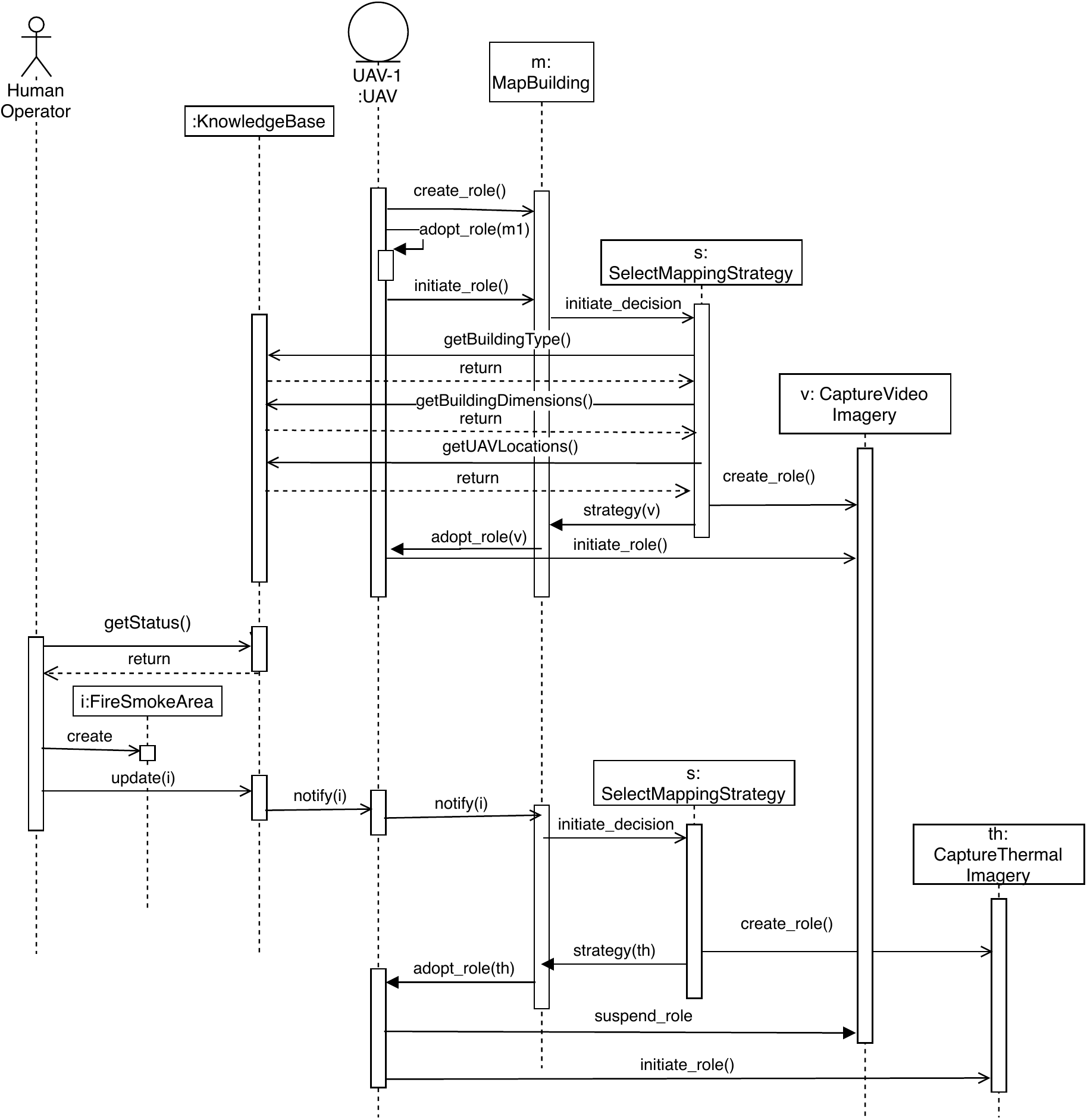}
    \caption*{b: Sequence diagram showing human-agent collaborations.}
    \caption{Multiple UAVs collaborate to create the 3D model of the building. When the human operator shares information about smoke observed at one end of the building, UAV-1 starts capturing thermal images to search for victims through the smoke.}
    \label{fig:sequence_fire_map}
    
\end{figure}

The probing questions enable us to explore this type of scenario. \textit{PQ11}, \textit{PQ12}, and \textit{PQ13} identify the required \texttt{AutonomousDecision}s and the required \texttt{Information} to create the 3D model of the building autonomously. \textit{PQ3} and \textit{PQ4} elicit human multi-UAV interaction points such as \textit{fire smoke detection by humans} while UAVs are engaged in mapping the building. \textit{PQ6} identifies potential flight adaptation patterns and roles assumed by the UAVs after receiving updated information about the smoke.  Answers to the probing questions lead us to construct the conceptual model and sequence diagram depicted in Figure~\ref{fig:sequence_fire_map}. 

\section{Threats to Validity}
\label{sec:threats}
There are several threats to validity for our approach. 
First, we have applied the probing questions retrospectively to construct the M1 models described in the fire-surveillance example; however, we answered the questions based on information gathered through a series of brainstorm meetings with firefighters. In the next phase of our work, we will further evaluate the questions in live requirements elicitation sessions.  Second, we developed our meta-model based on five use-cases primarily developed  by our own research group in collaboration with our local fire fighters. We then demonstrated its generalizability using an additional use case that we developed. Our approach needs to be evaluated on use-cases elicited from diverse groups of emergency responders.  Finally, our approach currently ends at the modeling stage. To fully evaluated the usefulness of the model and the probing questions, we need to implement and integrate the modeled interactions within our deployed system.  We are currently working towards developing the required infrastructure such as AI vision models, on-board analysis and reasoning framework to support autonomous capabilities of the UAVs, and will then evaluate the extent to which our approach produces a viable design for use with physical UAVs. 

\section{Related Work}
\label{sec:related}
The effectiveness of the HotL is highly dependent upon the human multi-agent interaction mechanisms built into the system as well as the flexibility of the autonomy models. 
To this end, several researchers have explored techniques for exposing the intent, actions, plans and associated rationales of an autonomous agent  \cite{chen2018situation}, while other researchers have explored ways to improve overall performance by dynamically adapting agents' autonomy levels based on the estimated cognitive workload of the human participants; however, they also observed that frequent changes in autonomy levels reduced situational awareness and forced operators to continually reevaluate the agents' behavior  \cite{heard2020sahrta}. 

Furthermore, systems that use AI techniques to support autonomy often lack adequate explanations of the autonomous behavior which can negatively impact achievement of mission goals \cite{stoica2017berkeley} and reduce trust in the system. Therefore, several of our PQs are  specifically designed to explore the explainable aspects of a HotL system. Guizzardi  argues that RE techniques can be applied in the design of AI systems, such as driverless cars and autonomous weapons, to ensure that they comply to ethical principles and codes \cite{guizzardi2020ethical}. Gamification is a popular technique for gathering and validating the requirements  of a cyber-physical system \cite{lombriser2016gamified}. Wiesner et al. 
\cite{wiesner2016supporting} engaged stakeholders in a simulated game under different operational conditions to discover the limitations of the existing requirements and to support the ideation of possible new services. Fischer also uses a multi-player mixed-reality game to generate requirements for interaction and coordination within rich and ‘messy’ real-world socio-technical settings \cite{fischer2014supporting}. However, Hyrynsalmi discusses limitations of gamification techniques \cite{hyrynsalmi2017dark}, for example, users focusing on winning the `game' instead of the challenges of interacting with the system \cite{knaving2013designing}. The gamification approach also requires a significant upfront development effort and proves insufficient to explore the unknown unknowns of the system.  Our work takes a more formal approach to elicit requirements using a concrete meta-model and PQs that focus on the human interaction aspects of the multi-agent HotL systems.

\section{Conclusion}
\label{sec:conclusion}
This paper describes the model-driven analysis and specification of human multi-agent interaction requirements for a human-on-the-loop system. The human multi-agent interaction types,  the proposed meta-model, and the structured probing questions assist in modeling and formally specifying the complex human multi-agent interactions. We have demonstrated its use through formally specifying human interaction and intervention points for two distinct scenarios in which multiple semi-autonomous UAVs are deployed in emergency response missions. Our future work will involve implementing and evaluating our models with first-responders with physical UAVs in outdoor field-tests.

\section{Acknowledgement}
The work described in this paper is partially funded by the USA National Science Foundation grant CNS-1931962 .

\balance
\bibliographystyle{IEEEtran}
\bibliography{modre_2020.bib}
\end{document}